\documentclass[prl,floatfix,amsmath,superscriptaddress,twocolumn,showpacs]{revtex4}

\usepackage{amssymb}
\usepackage{graphicx}
\usepackage{graphics}
\usepackage{amsmath}
\usepackage{amsthm}
\usepackage{amssymb}
\usepackage{url}
\usepackage{bm}
\usepackage{booktabs}
\usepackage{braket}
\usepackage{float}
\usepackage{bbold}
\usepackage{enumerate}
\usepackage{color}
\usepackage{hyperref}	
\hypersetup{
	colorlinks=true,				
	linkcolor=blue, 				
}

\def\sx{\sigma_x}

\renewcommand\Re{\operatorname{Re}}
\renewcommand\Im{\operatorname{Im}}

\newcommand{\bea}{\begin{eqnarray}}
\newcommand{\eea}{\end{eqnarray}}

\def\bi{\begin{itemize}}
\def\ei{\end{itemize}}
\def\bc{\begin{center}}
\def\ec{\end{center}}

\def\C{\hbox{$\mit I$\kern-.7em$\mit C$}}
\def\R{\hbox{$\mit I$\kern-.6em$\mit R$}}

\def\ket#1{|#1\rangle}
\newcommand{\one}{\mbox{$1 \hspace{-1.0mm}  {\bf l}$}}
\def\tr{\mathrm{tr}}
\def\ket#1{\left| #1\right>}

\newcommand{\abs}[1]{\left| #1 \right|} 

\newtheorem{theorem}{Theorem}

\newtheorem{lemma}[theorem]{Lemma}

\makeatletter
\renewcommand{\maketag@@@}[1]{\hbox{\m@th\normalsize\normalfont#1}}%
\makeatother 

\begin{document}

\author{K. Schwaiger}
\affiliation{Institute for Theoretical Physics, University of
Innsbruck, Innsbruck, Austria}

\author{D. Sauerwein}
\affiliation{Institute for Theoretical Physics, University of
Innsbruck, Innsbruck, Austria}

\author{ M. Cuquet}
\affiliation{Institute for Theoretical Physics, University of
Innsbruck, Innsbruck, Austria}

\author{J.I. de Vicente}
\affiliation{Departamento de Matem\'aticas, Universidad Carlos III de
Madrid, Legan\'es (Madrid), Spain}
\author{B. Kraus}
\affiliation{Institute for Theoretical Physics, University of
Innsbruck, Innsbruck, Austria}
\title{Operational multipartite entanglement measures}


\begin{abstract}
We introduce two operational entanglement measures which are applicable for arbitrary multipartite (pure or mixed) states. One of them characterizes the potentiality of a state to generate other states via local operations assisted by classical communication (LOCC) and the other the
simplicity of generating the state at hand. We show how these measures can be generalized to two classes of entanglement measures. Moreover, we compute the new measures for pure few-partite systems and use them to characterize the entanglement contained in a three-qubit state. We identify the GHZ- and the W-state as the most powerful pure three-qubit states regarding state manipulation.
\end{abstract}

\maketitle

Entanglement is of paramount importance in many fields of science. Due to its existence, applications such as teleportation, quantum computation, quantum simulation, and quantum error correction, to name a few, are feasible \cite{Nie}. Moreover, the application of entanglement theory in other fields of science, most prominently condensed matter physics, has opened new routes towards the understanding of quantum many-body systems \cite{Faz}. Due to its importance, an enormous effort has been made to qualify and quantify multipartite entanglement. Different entanglement classes have been identified and several entanglement measures have been introduced \cite{HoHo08}. Some of them originated from analyzing the potentiality of a state for a particular realization of an application, such as the localizable entanglement \cite{locEnt}, some others arose from the generalization of classical correlation measures, such as the generalization of the squashed entanglement \cite{ChWi,HoHo08}.

Despite these results, we are still far from completely understanding multipartite entanglement. The lack of knowledge stems on the one hand from the fact that the number of non-local parameters scales exponentially with the number of subsystems and, on the other hand, from the fact that the operations which are central in the investigation of entanglement, the local operations assisted by classical communication (LOCC), are notoriously difficult to be analyzed in general \cite{Chitambar}. The importance of LOCC in this context is due to the fact that LOCC corresponds to those operations which can be implemented without consuming entanglement. This implies that entanglement is the resource to overcome the restriction to LOCC and that the sole condition a function has to fulfill to be a valid entanglement measure is that it is non-increasing under LOCC \cite{HoHo08}, \footnote{Note that sometimes entanglement quantifiers which are non-increasing on average under LOCC have been considered (see \cite{HoHo08}).}. For the bipartite case a simple criterion for pure state transformations via LOCC has been presented \cite{Ni00}. These results do not only allow to identify the state $\ket{\Phi^+}\propto \sum_i \ket{ii}$ as the maximally entangled state, which can be transformed into any other bipartite state deterministically via LOCC, but also allowed to introduce new entanglement measures. Due to the existence of different SLOCC classes in the multipartite setting \cite{DuCi,Frank4qubits}, i.e. the existence of pairs of states which cannot even probabilistically be transformed locally into each other, there does not exist a single state which is {\it the} optimal resource to overcome LOCC. This is why a set of states, the maximally entangled set (MES) of $n$ subsystems, $MES_n$, has to be considered \cite{mes}. It is the minimal set of states from which any fully-entangled $n$-partite state can be obtained via LOCC.

In order to quantify entanglement, possible LOCC transformations among multipartite states have to be further investigated with the intention to identify new operational entanglement measures. This is precisely the aim of this paper. We introduce operational entanglement measures for multipartite states (pure or mixed) of arbitrary dimensions \footnote{Note that the fact that the measures are applicable for any dimension is crucial, as LOCC might increase the dimension of the considered Hilbert space.}. As we are going to show, the measures can be easily computed whenever all possible LOCC transformations are known, as in the case of pure states describing few-partite systems \cite{Ni00, turgut} \footnote{Note that the LOCC convertibility among mixed states is only partially known even in the case of bipartite systems \cite{hayden}.}. The operational character of the new measures allows to prove very easily that they are indeed non-increasing under LOCC and admits a generalization to two classes of entanglement measures.

We outline how to compute the new measures for bipartite pure states of arbitrary dimensions.  For pure three-qubit systems we derive explicit formulas for them and show that they, together with some well-known bipartite measures, allow us to completely characterize the entanglement contained in the state in an operational way. This characterization shows that the W- and GHZ-state are the most useful tripartite states regarding state manipulation.

Throughout this paper, $\sigma_x, \sigma_y, \sigma_z$ denote the Pauli operators and $\one$ the identity operator. When studying possible LOCC transformations we always consider representatives of Local Unitary (LU)-equivalence classes, as LUs do not alter the entanglement contained in a state and can obviously always be applied to a state. We say that a state $\ket{\Psi}$ can {\it reach} a state $\ket{\Phi}$ if there exists a LOCC protocol which transforms $\ket{\Psi}$ into $\ket{\Phi}$ (deterministically). In this case $\ket{\Phi}$ is {\it accessible} from $\ket{\Psi}$.

Let us now introduce the new operational entanglement measures. For a given state, $\ket{\Psi}$, we denote by $M_a(\ket{\Psi})$ the set of states which can be accessed via LOCC from $\ket{\Psi}$ and by $M_s(\ket{\Psi})$ the set of states which can reach $\ket{\Psi}$. The following two magnitudes occur then naturally in the context of possible LOCC transformations: the source volume, $V_s(\ket{\Psi})=\mu[M_s(\ket{\Psi})]$, which measures the amount of states that can be used to reach the state $\ket{\Psi}$ and the accessible volume, $V_a(\ket{\Psi})=\mu[M_a(\ket{\Psi})]$, which measures the amount of states that can be accessed by $\ket{\Psi}$ via LOCC. 
Here, $\mu$ denotes an arbitrary measure in the set of LU equivalence classes. The underlying idea is that if a state $\ket{\Psi}$ can be reached by many states, i.e. if $M_s(\ket{\Psi})$ is very large, then the state is not very powerful as any state in $M_s(\ket{\Psi})$ could be used for the same purpose and for possibly more applications. On the other hand, if the accessible set is very large the state is very valuable, as it can be used for any potential application of any state in $M_a(\ket{\Psi})$. 

Due to the operational meaning of $M_a$ and $M_s$, it is easy to construct now operational entanglement measures. In order to do so we first show that $M_a$ ($M_s$) can only become smaller (larger) under LOCC, respectively. Consider a state $\ket{\Psi}$ and any state $\ket{\Phi}_\Psi$ which is accessible from $\ket{\Psi}$ via LOCC. As any state in $M_s(\ket{\Psi})$ can first be transformed via LOCC into $\ket{\Psi}$ and then to $\ket{\Phi}_\Psi$, it is obvious that $M_s(\ket{\Phi}_\Psi)$ contains $M_s(\ket{\Psi})$. That $M_a(\ket{\Phi}_\Psi) \subseteq M_a(\ket{\Psi})$ can be easily verified noting that any state which can be reached from a state $\ket{\Phi}_\Psi$ can in particular be reached from a state that can reach $\ket{\Phi}_\Psi$. Hence, any properly normalized and rescaled measure of these sets is indeed an entanglement measure, i.e. it does not increase under LOCC. A possible choice would be $E_{a}(\ket{\Psi})=V_a(\ket{\Psi})/V_a^{sup}$ and $E_{s}(\ket{\Psi})=1-V_s(\ket{\Psi})/V_s^{sup}$, where $V_a^{sup}$ ($V_s^{sup}$) denote the maximally accessible (source) volume according to the measure $\mu$. Note that these operational entanglement measures are applicable to arbitrary multipartite systems of any dimension. Moreover, these are valid entanglement measures for mixed states.
Note further, that $M_s(\ket{\Psi})=\emptyset$ (implying that $V_s(\ket{\Psi})=0$) iff the state $\ket{\Psi}$ is in the MES, as these are the only states which cannot be reached by any other state \cite{mes}. We elaborate on how this measures can be computed in case of few-partite pure states below.

The notion of these entanglement measures can be generalized in the following way. Considering a $n$-partite state, $\ket{\Psi} \in \C^{d_1}\otimes \cdots \otimes \C^{d_n}$, one can also measure its entanglement by (i) the amount of $(n-k)$-partite entangled states one can reach from $\ket{\Psi}$, for $k\geq1$, or (ii) by the amount of reachable states in $\C^{d^\prime_1}\otimes \cdots \otimes \C^{d^\prime_n}$, where at least one of the local dimensions, $d_i^\prime$, is reduced. Similarly, one can generalize the notion of the source volume to a whole class of entanglement measures by relating not only elements of the same Hilbert space.

We are going to use now these quantities and the previously obtained results on possible LOCC transformations \cite{Ni00,turgut,KinTurg} in order to quantify the entanglement contained in few-body pure states. Let us start by considering the bipartite case. We consider without loss of generality two $d$ level systems. It is well known that a state $\ket{\Psi}$ can be transformed into a state $\ket{\Phi}$ via LOCC iff $\lambda_\Psi$ is majorized by $\lambda_\Phi$, i.e $\lambda_\Psi \preceq \lambda_\Phi$, where $\lambda_\Psi$ denotes the vector containing the eigenvalues of the single party reduced state of $\ket{\Psi}$ \cite{Ni00}. As the state is normalized, any vector $\lambda_\Psi$ belongs to a $d$-dimensional simplex. It has been shown that the set $S(y)=\{x \in \R^d |x \preceq y\}$ is the convex hull of $d !$ points obtained by permuting the components of $y$ \cite{Rado}. Hence, in this parameter space and using the Lebesgue measure, the source and accessible volume of a state $\ket{\Psi}$ are given by the volume of $S(\lambda_\Psi)$ and $A(\lambda_\Psi)=\{x \in \R^d |\lambda_\Psi \preceq x\}$ respectively (up to a constant factor, see \cite{SaCu15}). In \cite{SaCu15} we present closed formulas for $E_s$ and its generalization. Moreover, we present an algorithm to determine $E_a$ for arbitrary dimension and present explicit formulas for low dimensions, for which the new measures can be used to completely characterize the Schmidt coefficients.

Let us now present a complete characterization of entanglement of an arbitrary pure three-qubit state. In order to understand how the measures $E_a$ and $E_s$ are defined in this case we give a few remarks. First, we only consider the source and accessible volumes of genuinely entangled three-qubit states (i.\ e.\ we do not take into account biseparable states). Second, when we consider LOCC incomparable families of states such as the W and GHZ SLOCC classes, there is a freedom in choosing different measures $\mu_1$ and $\mu_2$ to compute the volumes for the different families without compromising the behaviour under LOCC of the entanglement measures. We exploit this freedom out of mathematical convenience. Last, even when considering LOCC comparable states, there exist states for which the corresponding source or accessible states are in manifolds of different dimensionality. Hence, if we use the same measure $\mu$ in both cases, this would assign a zero value to the accessible or source volumes of certain states even though they can indeed reach or be reached by other states, leading to a too coarse grained classification. Even though this would be a legitimate choice, we choose to use here a finer classification by choosing different measures to compute the volumes whenever the corresponding manifolds have different dimensions. Note that this choice, in contrast to the afore mentioned one, allows to compare the relative strength of states whose volumes have the same dimensionality. It should be clear however, that a state with e.g. a non-vanishing 4-dimensional accessible volume is infinitely more powerful than a state with a 3-dimensional accessible volume.

\begin{figure}[H]
   \centering
   \includegraphics[width=0.18\textwidth]{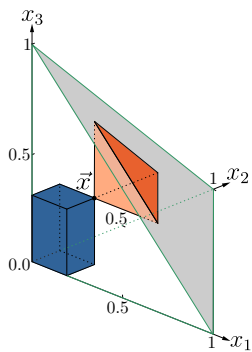}
   \caption{(color online). Any state $\ket{\Psi(\vec{x})}$ is uniquely represented by $\vec{x}$ in the interior of the simplex $\mathcal{S}_3$. The source (tetrahedron) and the accessible (cuboid) volume of $\ket{\Psi(\vec{x})}$ are depicted. The light surface corresponds to the states in the MES. Biseparable (fully separable) states, for which exactly one (at least two) $x_i$ is (are) zero,
   are represented by points on the white surface of $\mathcal{S}_3$, respectively.}
   \label{VolumesW}
   \end{figure}

Up to LUs, any state in the W-class can be written as \cite{mes,KinTurg}
\bea
\ket{\Psi(\vec{x})}\hspace{-0.1cm} = \hspace{-0.1cm}\sqrt{x_0}\ket{000}\hspace{-0.1cm}+\hspace{-0.1cm}\sqrt{x_1}\ket{100}\hspace{-0.1cm}+\hspace{-0.1cm}\sqrt{x_2}\ket{010}\hspace{-0.1cm}+\hspace{-0.1cm}\sqrt{x_3}\ket{001},
\eea
where $x_1, x_2, x_3 > 0$, $x_0 \geq 0$ and $\sum_{i=0}^3 x_i=1$. Note that any state $\ket{\Psi(\vec{x})}$ can be represented by the corresponding vector $\vec{x}= (x_1,x_2,x_3)$ within a three-dimensional simplex $\mathcal{S}_3$ (see Fig.\ \ref{VolumesW}) \cite{KinTurg}.
As shown in \cite{KinTurg}, $\ket{\Psi(\vec{x})}$ can be
transformed into $\ket{\Psi(\vec{y})}$ via LOCC iff $x_i \geq y_i, \forall i \in \{1,2,3\}$. Due to that, it can be easily verified that the MES in the W-class, the W-MES, is the set of states with $x_0 = 0$. These states cannot be obtained from any other state, but any state in the W-class can be obtained from some state with $x_0 = 0$ \cite{mes}.

Within the parameter space explained above, the accessible volume and the source volume of an arbitrary state $\ket{\Psi(\vec{x})}$ can be easily shown to be (see also Fig.\ \ref{VolumesW})
\bea
V_a(\ket{\Psi(\vec{x})}) = x_1 x_2 x_3, \label{eq_VaW} \ \
V_s(\ket{\Psi(\vec{x})}) = \frac{x_0^3}{6}.\label{eq_VsW}
\eea
As mentioned above, it follows already from their definition that the corresponding measures, $E_a(\ket{\Psi(\vec{x})}) = 27 V_a(\ket{\Psi(\vec{x})})$ and $E_s(\ket{\Psi(\vec{x})}) = 1 - 6 V_s(\ket{\Psi(\vec{x})})$, are entanglement measures.
A fact that can be particularly easily verified for the W-class using Eq.\ \eqref{eq_VaW} and that no $x_i$ can be increased via LOCC. Note that the W-state maximizes both new measures, with $E_a(\ket{W}) = 1$ and $E_s(\ket{W}) = 1$.
Hence, the W-state is the state that reaches the most other states deterministically via LOCC. It can therefore be regarded as the
most useful state in the W-class.

Let us now characterize the entanglement contained in a state in the W-class. Due to the simplicity of this class, only bipartite entanglement measures, e.g. the three bipartite entanglement between party $i$ and the remaining parties,
measured with e.g. the squared concurrence \cite{conc}, $C_i(\ket{\Psi(\vec{x})})= 4 x_i(1-x_i-x_0)$, are required to uniquely characterize the state (up to LUs). However, one could also employ the new measures and any of the bipartite measures for this purpose. In fact, as any three measures of the set $ \{C_1(\ket{\Psi}), C_2(\ket{\Psi}), C_3(\ket{\Psi}), E_a(\ket{\Psi}), E_s(\ket{\Psi})\}$ are independent, we have that a state in the W-class is uniquely determined by any three of these operational entanglement measures.

Note that for any state in the W-MES it holds that $V_s(\ket{\Psi(\vec{x})})=0$. Moreover, for these states we have $C_i(\ket{\Psi(\vec{x})}) = 4 x_i(1-x_i)$. Hence, they can be easily characterized by any two of these bipartite measures or by any bipartite measure and $E_a(\ket{\Psi(\vec{x})})$. Note that the characterization via operational entanglement measures of arbitrary states in the W-class presented here can be easily generalized to $n$-qubit systems \cite{SaCu15}.

Let us now investigate the more complicated GHZ-class.
A state in the GHZ-class can be written (up to LUs) as \cite{mes}
\bea \label{GHZ}
\ket{\Psi(\boldsymbol{g},z)} = g_x^1 \otimes g_x^2 \otimes g_x^3 P_z \ket{GHZ}, \eea
with $(g_x^i)^{\dagger} g_x^i = \frac{1}{2} \mathbb{1} + g_i \sx,$  $g_i \in [0,1/2)$ $\forall i$, $\boldsymbol{g} = (g_1, g_2, g_3)$, $ P_z = diag(z,1/z), \ z \in \mathbb{C}$ with $|z|\leq 1$. As shown in \cite{mes}, a state is in GHZ-MES iff $z= 1$ or $z=i$ and either non of the $g_i$'s vanishes or all of them vanish, corresponding to the GHZ-state.

In \cite{turgut} the necessary and sufficient conditions for the existence of a LOCC transformation from a state $\ket{\Psi(\boldsymbol{g},z)}$ to another state $\ket{\Psi(\boldsymbol{h},z')}$ were obtained. As shown there, the absolute value of $z$ can be changed by LOCC independently of the other parameters only if at least one of the parameters $g_i$ vanishes, in which case $|z|$ can be arbitrarily decreased. As in this case different LOCC transformations are possible, we treat the cases (A) $g_i \neq 0$ for all $i$ and (B) at least one of the parameters $g_i$ vanishes separately.

Let us first consider case (A). Expressing the conditions for the existence of a LOCC transformation \cite{turgut} from $\ket{\Psi(\boldsymbol{g},z)} $ to $\ket{\Psi(\boldsymbol{h},z')}$ (see Eq.\ \eqref{GHZ}) we obtain
\begin{enumerate}[(i)]
\item \label{Iq_GHZdet}$g_i \leq h_i \ \ \forall i,$
\item \label{Eq_GHZdet} $\displaystyle{\frac{g_1 g_2 g_3}{h_1 h_2 h_3} = \frac{\Re(z'^2)}{\abs{z'}^4+ 1} \frac{\abs{z}^4 + 1}{\Re(z^2)}= \frac{\Im(z'^2)}{\abs{z'}^4 - 1} \frac{\abs{z}^4 - 1}{\Im(z^2)}}$.
\end{enumerate}

Note that cond.\ \eqref{Eq_GHZdet} constitutes generically two independent equalities. However, in case the numerator and/or the denominator of one ratio vanishes, different conditions have to hold (see \cite{supp}).

We present now a characterization of the entanglement contained in an arbitrary state in the GHZ-class.

    \begin{figure}[H]
   \centering
   \includegraphics[width=0.2\textwidth]{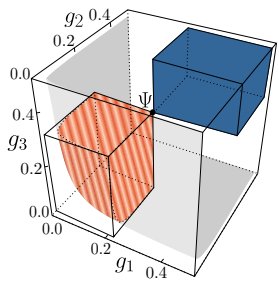}
   \caption{(color online). The source (shaded volume) and accessible (cuboid) volume of the state $\ket{\Psi}$ with parameters $g_1=0.22$, $g_2=0.26$, $g_3=0.32$ and $z= 0.1e^{i 2.68}$ are depicted. The states in the MES, which fulfill cond.\ \eqref{Eq_GHZdet} are on the light area.}
   \label{VolumesGHZ}
   \end{figure}

For this purpose we first consider states which are neither in GHZ-MES nor any $g_i$ vanishes, i.e. case (A) with $z \neq 1, i$. The other cases are treated below. The accessible and the source volume are given by (see \cite{supp})    \bea
    V_a(\ket{\Psi(\boldsymbol{g},z)})\hspace{-0.1cm}&=&\hspace{-0.1cm}\left(1/2 - g_1\right)\left(1/2 - g_2\right)\left(1/2 - g_3\right),\label{VAGHZ}\\
    V_s(\ket{\Psi(\boldsymbol{g},z)})\hspace{-0.1cm}&=&\hspace{-0.1cm} G (1\hspace{-0.1cm} +\hspace{-0.1cm} f_z  [\log\left( f_z \right)\label{VSGHZ}
    (1\hspace{-0.1cm}-\hspace{-0.1cm}\frac{1}{2} \log\left( f_z\right))\hspace{-0.1cm}-\hspace{-0.1cm}1 ]),
    \eea
    with $f_z =\frac{2 \abs{\Re(z^2)}}{1+\abs{z}^4}$ and $G= g_1 g_2 g_3$ (see Fig.\ \ref{VolumesGHZ}).

We show in \cite{supp} that $E_a(\ket{\Psi(\boldsymbol{g},z)})= 8 V_a(\ket{\Psi(\boldsymbol{g},z)})$ and $E_s(\ket{\Psi(\boldsymbol{g},z)})=1-8V_s(\ket{\Psi(\boldsymbol{g},z)})$ together with the three bipartite entanglement measures and one additional bit, that provides information about a specific state in the source set, uniquely determine the five parameters $z = r e^{i \phi}, g_1, g_2, g_3$ and therefore uniquely characterize the entanglement of the states up to complex conjugation (taken with respect to the computational basis) and LUs \footnote{Note that complex conjugation of a state corresponds to a redefinition of the complex unit and can therefore not alter any operational entanglement measure \cite{BKLU}, if one considers only the state of interest, $\ket{\Psi}$.} (see \cite{supp} for details). 

In \cite{supp} we show that the states where at least one $g_i$ vanishes can be treated similarly and that there the entanglement of the states is uniquely determined by the five operational entanglement measures.

It remains to consider the states in GHZ-MES, which constitute a three parameter family. In this case only cond.\ \eqref{Iq_GHZdet} and the first equation in cond.\ \eqref{Eq_GHZdet} have to be fulfilled, which implies that only one parameter of the accessible states is fixed via cond.\ \eqref{Eq_GHZdet}. Hence, for $\ket{\Psi_{MES}}$, a state in $MES_3$, we obtain the four dimensional accessible volume \begin{small}
    \bea
    V_a(\ket{\Psi_{MES}})=  \int_{g_1}^{1/2}\hspace{-0.2cm}\int_{g_2}^{1/2}\hspace{-0.2cm}\int_{\abs{g_3}}^{1/2} \hspace{-0.2cm}\int_{\sqrt{\frac{H}{G} - \sqrt{\left(\frac{H}{G}\right)^2-1}}}^1 dr dh_3 dh_2 dh_1, \label{VaMES}
    \eea \end{small}
    with $G=g_1 g_2 \abs{g_3}, \ H = h_1 h_2 h_3$. Putting the lower limits in Eq.\ \eqref{VaMES} to zero we find $V_a^{sup}=1/8$. As the GHZ-state fulfills $g_i = 0 \ \forall i$ we have $E_a(\ket{GHZ})=1$ and therefore the GHZ-state is the state which reaches the most other states deterministically. The entanglement of a state in GHZ-MES, for which $V_s(\ket{\Psi_{MES}})=0$, can be similarly easily characterized as it was possible in the W-class (see \cite{supp}).

In summary, we have introduced two novel classes of operational entanglement measures, which are applicable to arbitrary multipartite pure or mixed states.
We then demonstrated how these measures can be computed for the simplest pure multipartite case (three qubits) and showed that they can be used to completely characterize the entanglement contained in a three-qubit state. In \cite{SaCu15} the new measures and its generalizations are determined for the bipartite setting of low dimension and the four-qubit case. It would be interesting to develop further the analysis of LOCC convertibility among mixed states in order to compute our measures in this case. Besides that and further extensions of this approach (e.\ g.\ approximate LOCC transformations, multi-copy case), it would also be appealing to connect our measures with different quantum information protocols and condensed-matter phenomena, which we leave for future research.

\begin{acknowledgments}
This research was funded by the Austrian Science Fund (FWF) Grant No. Y535-N16 and the Spanish MINECO
through grants MTM 2010-21186-C02-02, MTM 2011-26912 and MTM2014-54692 and the CAM regional research consortium QUITEMAD+CM S2013/ICE-2801.
\end{acknowledgments}

\part{Supplemental Material}

In this Supplemental Material we derive all formulas for the source and accessible volume for the states in the GHZ-class and prove the unique characterization of the entanglement of these states with our set of measures.

\section{The source and accessible volume of states in the GHZ-class}
  In this section we derive the formulas for the accessible and source volume of states in the GHZ-class. We will distinguish subsequently generic states, defined as those states that are not in $MES_3$ and for which no $g_i$ parameter vanishes, from non-generic ones. The reason for that is, as mentioned in the main text, that generic and non-generic states allow for different LOCC transformations. This fact is also reflected in the dimensionality of the two measures, that varies for generic and non-generic states. For generic states the source and accessible volume are always three-dimensional, whereas for non-generic states the volumes can be up to four-dimensional. For example the accessible volume of states in the MES is four-dimensional and the source volume is equal to zero.\\
  In order to improve readability we summarize here the necessary and sufficient conditions for state transformations from \cite{Aturgut}, where one has to distinguish the following cases.
  \begin{enumerate}[I.]
  \item For states with non-vanishing $g_i$, $h_i$ parameters the necessary and sufficient conditions for the existence of a LOCC transformation from $\ket{\Psi(\boldsymbol{g},z)} $ to $\ket{\Psi(\boldsymbol{h},z')}$ are given by
\begin{enumerate}[(i)]
\item \label{AIq_GHZdet}$g_i \leq h_i \ \ \forall i,$
\item \label{AEq_GHZdet} $\displaystyle{\frac{g_1 g_2 g_3}{h_1 h_2 h_3} = \frac{\Re(z'^2)}{\abs{z'}^4+ 1} \frac{\abs{z}^4 + 1}{\Re(z^2)}= \frac{\Im(z'^2)}{\abs{z'}^4 - 1} \frac{\abs{z}^4 - 1}{\Im(z^2)}}$.
\end{enumerate}
As stated in the main text, cond.\ \eqref{AEq_GHZdet} constitutes generically two independent equalities. However, we have to distinguish the following different cases when the numerator and/or the denominator of one ratio vanishes.
\begin{itemize}
\item In case both the numerator and the denominator of one ratio vanish, one has to ignore the corresponding equality and only the remaining equality has to be satisfied. 
\item If only the denominator or the nominator (but not both) vanishes, additional conditions have to be satisfied. These are: if $\Re(z^2)=0$ also $\Re(z'^2)$ has to vanish; if $\Im(z^2)=0$ either $\Im(z'^2)=0$ or $\abs{z}=1$, and if $\abs{z}=1$ then either $\abs{z'}=1$ or also $\Im(z^2)=0$ has to be fulfilled. Note that a state with $\Im(z^2)=0$ and $\abs{z}=1$ is a state in the MES.
\end{itemize} 
\item For states with vanishing $g_i$, $h_i$ parameters we have to distinguish the following two cases.
    \begin{enumerate}[1.]
    \item The necessary and sufficient conditions for the existence of a LOCC transformation from $\ket{\Psi(\boldsymbol{g},z)} $ (with $\boldsymbol{g}$ arbitrary) to a state with at least one vanishing parameter, i.e. $\ket{\Psi(\boldsymbol{h}: \exists i : h_i = 0,z')}$, read
\begin{enumerate}[(i)]
\item \label{h10} $g_i \leq h_i \ \forall i$,
\item \label{h10r} $r \geq r'$.
\end{enumerate}
\item The necessary and sufficient conditions for the existence of a LOCC transformation from a state with vanishing parameters, i.e. $\ket{\Psi(\boldsymbol{g}: \exists i : g_i = 0,z')}$, to a state with non-vanishing parameters, i.e. $\ket{\Psi(\boldsymbol{h}: h_i \neq 0 \forall i, z')}$, are given by
\begin{enumerate}[(i)]
\item $g_i \leq h_i \ \forall i,$ \label{r11}
\item $r = 1$, \label{r12}
\item $z' = r' e^{i \phi'},$ with $\phi' = \pi/4, 3 \pi/4$ and $r'$ arbitrary. \label{r13}
\end{enumerate}
    \end{enumerate}
  \end{enumerate}
In the subsequent subsections we use these criteria to determine the source and accessible volume.

\subsection{Generic states in the GHZ-class}
Let us start with the determination of the two measures for generic states in the GHZ-class. The expressions for the volumes of generic states, i.e.
    $V_a(\ket{\Psi(\boldsymbol{g},z)}), V_s(\ket{\Psi(\boldsymbol{g},z)})$ are given in the main text in Eqs.\ (5)-(6). In order to improve readability we restate the expressions here,
      \bea
   \hspace*{-0.3cm} V_a(\ket{\Psi(\boldsymbol{g},z)})\hspace{-0.1cm}&=&\hspace{-0.1cm} \left(1/2 - g_1\right)\left(1/2 - g_2\right)\left(1/2 - g_3\right), \label{VAGHZ1}\\
   \hspace*{-0.3cm} V_s(\ket{\Psi(\boldsymbol{g},z)})\hspace{-0.1cm}&=&\hspace{-0.2cm}  G (1\hspace{-0.1cm} +\hspace{-0.1cm} f_z  [\log\left( f_z \right) \label{VSGHZ1}
    (1\hspace{-0.1cm}-\hspace{-0.1cm}\frac{1}{2} \log\left( f_z\right))\hspace{-0.1cm}-\hspace{-0.1cm}1 ]),
    \eea
    with $f_z =\frac{2 \abs{\Re(z^2)}}{1+\abs{z}^4}$ and $G = g_1 g_2 g_3$ and $0 \log(0)=0$.\\
    We start with the derivation of the accessible volume
    of $\ket{\Psi(\boldsymbol{g},z)}$, using conditions \eqref{AIq_GHZdet} and \eqref{AEq_GHZdet}. Solving the two equations in cond.\ \eqref{AEq_GHZdet} determines the parameter $z'
    = r' e^{i \phi'}$ of $\ket{\Psi(\boldsymbol{h},z')}$. We find that the parameter $z'$ of all states in $M_a(\ket{\Psi(\boldsymbol{g},z)})$ has to fulfill
    \begin{small}
    \bea
    \tan(2 \phi') = \frac{r'^4 - 1}{r'^4 + 1}\frac{b_z}{a_z}, \ \
    r'^4_{\pm}=\frac{n}{d} \hspace{-0.05cm}\pm \hspace{-0.1cm} \sqrt{\left(\frac{n}{d}\right)^2\hspace{-0.1cm}-1}, \label{rVR}
    \eea
    \end{small}
    where we used the notation $a_z = \frac{2 \Re(z^2)}{\abs{z}^4 + 1}, \ b_z = \frac{2 \Im(z^2)}{\abs{z}^4 - 1},  \ n= b_z^2- a_z^2 + \frac{2 H^2}{G^2}, \ H = h_1 h_2 h_3, \ d= a_z^2 + b_z^2$. Note that from Eq.\ \eqref{rVR} we would get two possible solutions for $\phi' \in [0,\pi]$, namely $\phi'$ and $\phi' + \pi/2$, for every possible value of $r'$. However, from cond.\ \eqref{AEq_GHZdet} and the fact that $G/H$ is always positive it follows that the sign of $\Re(z^2)$ ($\Im(z^2)$) has to be the same as the sign of $\Re(z'^2)$ ($\Im(z'^2)$). Thus, only one of the two solutions for $\phi'$ is valid.
    Furthermore, the absolute value $r'$ of the complex number $z'$ has to be real, hence the expression under the square root in Eq.\ \eqref{rVR} has
    to be greater than or equal to zero. Using that $\abs{a_z} \leq 1 $ $\forall z$, one can easily see that this is satisfied by all $h_i$'s which fulfill the necessary cond.\ \eqref{AIq_GHZdet}, i.e. $h_i \geq g_i \ \forall i$. We obtain two solutions for $z'$, namely $z'_+ = z'$ and $z'_- = 1/z'$, for each possible combination of parameters $h_i$, fulfilling
    the inequalities in cond.\ \eqref{AIq_GHZdet}. However, a state with parameters $h_i$ and $z'_+$ is LU-equivalent to a state with parameters
    $h_i$ and $z'_-$, i.e. $\ket{\Psi(\boldsymbol{h},z'_+)} \stackrel{LU}{\simeq} \ket{\Psi(\boldsymbol{h},z'_-)}$ \cite{Aturgut}. Thus, we count only one solution for $z'$. Hence, the volume of
    states reachable via LOCC from $\ket{\Psi(\boldsymbol{g},z)}$ is computed as
    \bea
    V_a(\ket{\Psi(\boldsymbol{g},z)}) =  \int_{g_1}^{1/2}\int_{g_2}^{1/2}\int_{g_3}^{1/2} dh_3 dh_2 dh_1.\label{Vaint}
    \eea
    The result of this simple integral is given in Eq.\ \eqref{VAGHZ1}. It is easy to see that the supremum of the accessible volume for generic states is given by $V_a^{sup} = 1/8$, where one simply sets the lower limits in all three integrals to zero. Note that for the special cases where cond.\ \eqref{AEq_GHZdet} constitutes a single equation, we obtain the same accessible volume as in Eq.\ \eqref{Vaint}, as can be easily seen taking into account that also in this case two parameters are determined. That is for states with either $\Re(z^2) = \Re(z'^2) = 0$, $\Im(z^2)=\Im(z'^2)=0$ or
    $\abs{z} = \abs{z'}=1$ (excluding the states in the MES) the accessible volume is also given by Eq.\ \eqref{VAGHZ1}.\\
    Now we can move on to the source volume. Let us first consider the case where $\Re(z^2) \neq 0$ and $\Re(z'^2)\neq 0$. To determine $V_s(\ket{\Psi(\boldsymbol{g},z)})$ we need to interchange the parameters in cond.\ \eqref{AEq_GHZdet} and cond.\ \eqref{AIq_GHZdet}, i.e. $h_i$ is interchanged with $g_i$ and  $z$ is interchanged with $z'$. Then the parameter $z'$ of states in $M_s(\ket{\Psi(\boldsymbol{g},z)})$ has to fulfill
   \begin{small}
    \bea
    \tan(2\phi')= \frac{r'^4-1}{r'^4+1}\frac{b_z}{a_z}, \ \
    r'^4_{\pm}=\frac{n}{d} \hspace{-0.05cm}\pm \hspace{-0.1cm} \sqrt{\left(\frac{n}{d}\right)^2\hspace{-0.1cm}- 1}, \label{rVS}
    \eea
    \end{small}
    where we used the same notation as before and consider only one solution for $\phi'$ as explained above.
    Again the expression under the square root in Eq.\ \eqref{rVS} has to be greater than or equal to zero. In contrast to before, not all values of $h_i$ fulfilling cond.\ \eqref{AIq_GHZdet}, i.e. $h_i \leq g_i \ \forall i$, are allowed. Thus, we cannot set the lower limits of the integral over the $h_i$ to zero, but we have to find the correct lower limits via the condition $\left(\frac{n}{d}\right)^2- 1 \geq 0$. Simplifying this inequality shows that the states corresponding to the lower limits of the integral are states in the MES, i.e. $z' = 1,i$, which obtain $\ket{\Psi(\boldsymbol{g},z)}$ deterministically. In Fig.\ 3 in the main text these states lie on the intersection of the gray area with the red cuboid. Furthermore, the $h_i$ parameters of these states in the MES satisfy the following equation
    \bea
    h_1 h_2 h_3 = \frac{2 \Re(z^2) g_1 g_2 g_3}{1+\abs{z}^4} \frac{1}{\Re(z'^2)} \equiv \abs{a_z} G ,
    \eea
    i.e. cond.\ \eqref{AEq_GHZdet}. Thus, the source volume of states that can be converted deterministically into a given state $\ket{\Psi(\boldsymbol{g},z)}$ via LOCC is given by
    \bea
    V_s(\ket{\Psi(\boldsymbol{g},z)}) = \int_{\frac{\abs{a_z}G}{g_2 g_3}}^{g_1} \int_{\frac{\abs{a_z}G}{h_1 g_3}}^{g_2} \int_{\frac{\abs{a_z}G}{h_1 h_2}}^{g_3} dh_3 dh_2 dh_1.
    \eea
 The result of the source volume is given in Eq.\ \eqref{VSGHZ1} (with $f_z = \abs{a_z}$). It is again easy to see that the supremum of the source volume for generic states is given by $V_s^{sup} = 1/8$, where one sets the lower limits of the three integrals to zero and the upper limits to $1/2$.\\
    It remains to compute the source volumes for states with $\Re(z^2)=0$ (and hence also $\Re(z'^2)=0$).
    In this case cond.\ \eqref{AEq_GHZdet} constitutes a single equation. It can be shown that this equation is fulfilled for all $h_i \leq g_i$.  Hence, the source volume is given by
    \bea
   \hspace*{-0.3cm} V_s(\ket{\textstyle{\Psi(\boldsymbol{g},z:}\scriptstyle{\Re(z^2)=0}\textstyle{)}}\textstyle{ = \hspace{-0.1cm}\int_0^{g_1}\hspace{-0.1cm}\int_0^{g_2}\hspace{-0.1cm}\int_0^{g_3} dh_3 dh_2 dh_1 = G,}
    \eea
    with $G= g_1 g_2 g_3$.
    Note that one would obtain the same result using Eq.\ \eqref{VSGHZ1}, as in this case $f_z = 0$. Hence, also the supremum of the source volume stays the same, i.e. $V_s^{sup} = 1/8$.

\subsection{Non-generic states in the GHZ-class}
\label{CLU}
As mentioned in the main text, non-generic states in the GHZ-class have to be treated separately.
For states in $MES_3$ the source volume is equal to zero and the integral for the accessible volume is given in the main text in Eq.\ (8). It is a four-dimensional volume, as only the parameter $\phi$ is determined by cond. (\ref{AEq_GHZdet}).

Here, we derive the accessible and source volume of non-generic states in the GHZ-class where at least one of the $g_i$ parameters is equal to zero. These states have to fulfill different necessary and sufficient conditions for deterministic LOCC conversions than generic states, especially cond.\ \eqref{AEq_GHZdet} does not have to be fulfilled.
Note that these states are LU-equivalent to their complex conjugate, as we can choose the $z$ parameter to be real. One can simply apply a local unitary on the qubit with the vanishing $g_i$ parameter, leading to a vanishing phase of the parameter $z$, i.e. $z=r \in
(0,1]$ \footnote{Note that a state is LU-equivalent to its complex conjugate iff either $|z|=1$, $z \in \R$, $z \in i\R$, or at least one of the $g_i$'s vanishes. This can be easily verified using the fact that the eigenvalues of the reduced states of a state and its complex conjugate coincide and the fact that two such states are either LU-equivalent or LOCC incomparable \cite{Be}. Using then the necessary and sufficient condition for LOCC transformation presented in \cite{Aturgut} leads to the result.}. Thus, for states with vanishing parameters we will write in the following simply $r$ instead of $z$. \\
Let us first consider the source volumes of states with vanishing parameters, i.e. $g_i = 0$ for some $i \in \{1,2,3\}$.
To determine $V_s(\ket{\Psi(\boldsymbol{g}: \exists i : g_i = 0,r)})$ we again have to interchange the parameters in cond.\ \eqref{h10} and cond.\ \eqref{h10r}, i.e. $h_i$ is interchanged with $g_i$ and $r'$ is interchanged with $r$. From cond.\ \eqref{h10} it follows that if $g_i = 0 $ for some $i \in \{1,2,3\}$ then also $h_i = 0$ for the same $i$. Furthermore, it follows from cond.\ \eqref{h10r} that if $r=1$ then also $r'=1$. Thus, in order to compute the source volume of $\ket{\Psi(\boldsymbol{g},z)}$ we have to distinguish different cases, depending on the number of vanishing parameters and on $r$ being or not being equal to one. Here, we consider first the case $r \neq 1$.
\begin{enumerate}[(1)]
\item If exactly one of the $g_i$ parameters is equal to zero ($g_i = 0$ for exactly one $i \in \{1,2,3\}$), the source volume is given by
    \bea V_s(\ket{\Psi(\boldsymbol{g},r)}) &=& \hspace{-0.1cm} \int_{0}^{g_j} \hspace{-0.1cm} \int_{0}^{g_k} \hspace{-0.1cm} \int_{r}^1  dr' dh_k dh_j \nonumber  \\ &=& g_j g_k (1-r), \eea for
    $j\neq k \neq i$. The supremum of the source volume can be easily computed putting the upper limits of $g_j, g_k$ to $1/2$ and the lower limit of $r$ to zero leading to $V_s^{sup} = 1/4$.
\item If exactly two of the $g_i$ parameters are equal to zero ($g_i = g_j = 0$ for $i \neq j, \  i,j \in \{1,2,3\}$), the source volume reads
\bea V_s (\ket{\Psi(\boldsymbol{g},r)}) = \int_{0}^{g_k} \int_{r}^1 dr' dh_k  = g_k (1-r),\eea for $j\neq k \neq i$. The supremum of the source volume is computed in the same way as in case (1) leading to $V_s^{sup} = 1/2$.
\item If all of the $g_i$ parameters are equal to zero ($g_1 = g_2 = g_3 = 0$), the source volume reduces to the following one-dimensional integral
\bea V_s (\ket{\Psi(\boldsymbol{g},r)}) = \int_{r}^1 dr' =(1-r), \eea with $V_s^{sup} = 1$.
\end{enumerate}
Thus, depending on the number of $g_i$ parameters being equal to zero the states either have a three-, two- or one-dimensional source volume.

Let us now consider the remaining case $r=1$. Then also $r'$ has to be equal to one and therefore, has to have a fixed value for the states in the source set. Thus, the dimension of the source volume decreases by one. In more detail we find for $g_i = 0$ for exactly one $i \in \{1,2,3\}$ and $r=1$, that $V_s(\ket{\Psi(\boldsymbol{g},r: r=1)}) =  \int_{0}^{g_j}
\int_{0}^{g_k}  dh_k dh_j = g_j g_k$, for $g_i = g_j = 0$ for $i \neq j, \ i,j \in \{1,2,3\}$ and $r=1$, that $V_s(\ket{\Psi(\boldsymbol{g},r: r=1)}) =  \int_{0}^{g_k} dh_k = g_k$ and for $g_1 = g_2 = g_3 = 0$ and $r=1$ the
state is the GHZ-state, which has a vanishing source volume, i.e. $V_s(\ket{GHZ}) = 0$. \\
Now we derive the accessible volume of states with vanishing $g_i$ parameters, i.e. $g_i = 0$ for some $i \in \{1, 2, 3\}$. We again have to distinguish different cases, depending on the number of vanishing $g_i$'s and on $r$ being or not being equal to one. Let us start also here with the determination of the accessible volume of states $\ket{\Psi(\boldsymbol{g},r)}$ with $r\neq 1$. Recall that in this case conditions \eqref{h10} and \eqref{h10r} have to hold. Hence, we have:
\begin{enumerate}[(1)]
\item If exactly one of the $g_i$ parameters is equal to zero ($g_i = 0$ for exactly one $i \in \{1,2,3\}$), then as $r\neq1$ the accessible states have to fullfill $h_i = 0$ and hence, the accessible volume is given by
\bea V_a(\ket{\Psi(\boldsymbol{g},r)}) &=& \int_{g_j}^{1/2} \int_{g_k}^{1/2} \int_0^r  dr' dh_k dh_j \nonumber \\ &=&(1/2 - g_j)(1/2-g_k) r,\eea for $j\neq k \neq i$, with $V_a^{sup} = 1/4$.
\item If exactly two of the $g_i$ parameters are equal to zero ($g_i = g_j =0$ for $i \neq j, \ i,j  \in \{1,2,3\}$), then only states with either $h_i = 0$ and $h_j \in [0,1/2)$ or $h_j = 0$ and $h_i \in [0,1/2)$ can be reached and the accessible volume reads
\bea V_a(\ket{\Psi(\boldsymbol{g},r)}) &=& 2 \int_{0}^{1/2} \int_{g_k}^{1/2} \int_0^r  dr' dh_k dh_j \nonumber \\ &=&(1/2-g_k) r, \eea for $j\neq k \neq i$, with $V_a^{sup} = 1/2$. Note that the factor of 2 is due to the fact, that we can choose either $h_i$ to be equal to zero and $h_j \in [0,1/2)$ or the other way around.
\item If all of the $g_i$ parameters are equal to zero ($g_1 = g_2 = g_3 = 0$), then the accessible states have to have at least one vanishing $h_i$, i.e. $h_i=0$ for some $i \in \{1,2,3\}$ and hence the accessible volume is given by
\bea V_a(\ket{\Psi(\boldsymbol{g},r)}) &=& 3 \int_{0}^{1/2} \int_{0}^{1/2} \int_0^r dr' dh_3 dh_2   \nonumber \\ &=& 3/4 r,\eea with $V_a^{sup} = 3/4$.
The factor 3 in front of the integral is again due to the fact that we can choose any $h_i$ for some $i \in \{1,2,3\}$ to be equal to zero, while the other two parameters can have any value between 0 and 1/2.
\end{enumerate}

Let us now discuss the accessible volume of states with vanishing parameters and $r=1$. These states fulfill cond.\ \eqref{r12} and can therefore reach all states $\ket{\Psi(\boldsymbol{h},z')}$ fulfilling cond.\ \eqref{r11} and cond.\ \eqref{r13}.
Hence, the accessible volume of $\ket{\Psi(\boldsymbol{g},r: r=1)}$ where at least one of the $g_i$'s is equal to zero is given by
\bea
V_a(\ket{\Psi(\boldsymbol{g},r\hspace{-0.1cm}: r\hspace{-0.1cm}=\hspace{-0.1cm}1)}) &\hspace{-0.1cm}=\hspace{-0.1cm}& 2 \hspace{-0.1cm}\int_{g_1}^{1/2}\hspace{-0.2cm}\int_{g_2}^{1/2} \hspace{-0.2cm}\int_{g_3}^{1/2} \hspace{-0.2cm}\int_0^1 dr' dh_3 dh_2 dh_1  \hspace{-0.1cm} \nonumber \\&=& 2 (1/2\hspace{-0.1cm}-\hspace{-0.1cm}g_1)(1/2\hspace{-0.1cm}-\hspace{-0.1cm}g_2)(1/2\hspace{-0.1cm}-\hspace{-0.1cm}g_3). \label{Var1}
\eea
Note that the factor of 2 in front of the integral is due to the fact, that $\phi' =\pi/4$ or $\phi'=3 \pi/4$ (see cond.\ \eqref{r13}).
The supremum of the accessible volume is then given by $V_a^{sup} = 1/4$.
Note that these states, similar to the states in the MES, have a four-dimensional accessible volume, which is a subset of the accessible volume of the
GHZ-state. The accessible volume of the GHZ-state is given by Eq.\ \eqref{Var1} with $g_1 =g_2 =g_3=0$, i.e. $V_a(\ket{GHZ})=1/4$. Furthermore, the GHZ-state is the only state in the MES, that reaches states with a vanishing $h_i$ parameter deterministically. This can be easily seen considering cond.\ \eqref{h10}. The condition implies that we can only increase the $g_i$'s via LOCC. Hence, the GHZ-state is the only state in the MES, that can reach a state with a vanishing parameter, as all other states in the MES have non-vanishing $g_i$ parameters.\\
Let us summarize here the new measures we have found for states with vanishing $g_i$'s. We start with states fulfilling $r \neq 1$.
\begin{enumerate}[(a)]
\item For a state $\ket{\Psi(\boldsymbol{g},r)}$ with one vanishing parameter, i.e. $g_i = 0$ for exactly one $i \in \{1,2,3\}$, we find
\bea
E_a(\ket{\Psi(\boldsymbol{g},r)}) &=& 4 (1/2- g_j)(1/2-g_k)r, \\
E_s(\ket{\Psi(\boldsymbol{g},r)}) &=& 1 - 4 g_j g_k (1-r),
\eea
with $i \neq j \neq k$ and a three-dimensional accessible and source volume.
\item For a state $\ket{\Psi(\boldsymbol{g},r)}$ with two vanishing parameters, i.e. $g_i = g_j = 0$ for $i \neq j, \ i,j \in \{1,2,3\}$, the new measures are given by
\bea
E_a(\ket{\Psi(\boldsymbol{g},r)}) &=& 2 (1/2-g_k)r, \\
E_s(\ket{\Psi(\boldsymbol{g},r)}) &=& 1 - 2 g_k (1-r),
\eea
with $i \neq j \neq k$, a three-dimensional accessible volume and a two-dimensional source volume.
\item For a state $\ket{\Psi(\boldsymbol{g},r)}$ with three vanishing parameters, i.e. $g_1 = g_2 =g_3 = 0$, the measures in terms of the fourth parameter $r$ read
\bea
E_a(\ket{\Psi(\boldsymbol{g},r)}) &=& r, \\
E_s(\ket{\Psi(\boldsymbol{g},r)}) &=& r,
\eea
with a three-dimensional accessible volume and a one-dimensional source volume.
\end{enumerate}
For states with vanishing parameters and $r=1$ we find the following new measures.
\begin{enumerate}[(a)]
\item For a state $\ket{\Psi(\boldsymbol{g},r: r=1)}$ with one vanishing parameter, i.e. $g_i = 0$  for exactly one $i \in \{1,2,3\}$ the measures are given by
\bea
E_a(\ket{\Psi(\boldsymbol{g},r: r=1)}) &=& 4 (1/2- g_j)(1/2-g_k), \\
E_s(\ket{\Psi(\boldsymbol{g},r: r=1)}) &=& 1 - 4 g_j g_k,
\eea
with $i \neq j \neq k$, a four-dimensional accessible volume and a two-dimensional source volume.
\item For a state $\ket{\Psi(\boldsymbol{g},r: r=1)}$ with two vanishing parameters, i.e. $g_i = g_j = 0$ for $i \neq j, \ i,j \in \{1,2,3\}$ we get
\bea
E_a(\ket{\Psi(\boldsymbol{g},r: r=1)}) &=& 2 (1/2-g_k), \\
E_s(\ket{\Psi(\boldsymbol{g},r: r=1)}) &=& 1 - 2 g_k,
\eea
with $i \neq j \neq k$, a four-dimensional accessible volume and a one-dimensional source volume.
\item The state $\ket{\Psi(\boldsymbol{g},r: r=1)}$ with three vanishing parameters, i.e. $g_1 = g_2 =g_3 = 0$ is the GHZ-state, with
\bea
E_a(\ket{GHZ}) &=& 1, \label{VaGHZ} \\
E_s(\ket{GHZ}) &=& 1,
\eea
a four-dimensional accessible volume and an empty source volume.
\end{enumerate}

\section{Unique characterization of the entanglement of states in the GHZ-class}
In this section we present the details of the proof that we can uniquely characterize the entanglement contained in a state in the GHZ-class with the help of the two new operational entanglement measures, three bipartite entanglement measures and an additional bit value $b$. The reason for this additional bit is that at most two states $\ket{\Psi(\boldsymbol{g},z_1)}$ and $\ket{\Psi(\boldsymbol{g},z_2)}$ are compatible with the five entanglement measures. Comparing the bipartite entanglement of two specific states that are in the source set of $\ket{\Psi(\boldsymbol{g},z_1)}$ and $\ket{\Psi(\boldsymbol{g},z_2)}$, respectively, and are both in the MES, allows to uniquely identify the state of interest. That is, the entanglement of $\ket{\Psi(\boldsymbol{g},z)}$ can be uniquely characterized (up to complex conjugation \footnote{As mentioned in the main text complex conjugation of a state corresponds to a redefinition of the complex unit and can therefore not alter any operational entanglement measure \cite{BKLUpaper}, if one considers only the state of interest, $\ket{\Psi}$. However, in case some additional state is considered, $\ket{\Psi}$ can be distinguished from its complex conjugate. In fact, similarly to the bit defined here, one could define a bit which distinguishes the two states.}) by the bit value b and the following set of operational entanglement measures \footnote{ Note that one could easily distinguish $\ket{\Psi}$ from its complex conjugate if not only $\ket{\Psi}$, but an additional state is considered.},
\bea
 \{C_1(\ket{\Psi}), C_2(\ket{\Psi}), C_3(\ket{\Psi}), E_s(\ket{\Psi}), E_a(\ket{\Psi})\}. \label{Measures}
\eea
We will again distinguish between generic and non-generic states in the GHZ-class, as non-generic states allow for different LOCC operations and furthermore, the dimensionality of the measures changes, as shown in the previous section.

\subsection{Uniqueness of generic states in the GHZ-class}
Here we want to show in more detail that the five measures and the bit value mentioned above uniquely define the entanglement of generic states $\ket{\Psi(\boldsymbol{g},z)}$ in the GHZ-class. The outline of the proof is as follows.
We first express the three parameters $g_2, g_3, \cos(2\phi)$ in terms of the three bipartite measures $C_i(\ket{\Psi(\boldsymbol{g},z)})$, $g_1$ and $r$. Then we insert the expressions for $g_2$ and $g_3$ (they do not depend on $r$) in the expression of $E_a(\ket{\Psi(\boldsymbol{g},z)})$. Solving this equation is equivalent to finding the roots of a function depending solely on $g_1$. As this function is strictly decreasing in $g_1$, there exists a unique solution for $g_1$. Thus, also $g_2$ and $g_3$ are uniquely defined, as they depend solely on $g_1$. Inserting the solutions for the $g_i$'s in the expression for $\cos(2 \phi)$ (which now solely depends on $r$) and using $E_s(\ket{\Psi(\boldsymbol{g},z)})$, we obtain an equation for $r$. One can show that this equation has at most two solutions for $r$ and corresponding solutions for $\phi$. This leads to at most two generic states in the GHZ-class compatible with the five entanglement measures in Eq.\ \eqref{Measures}. We distinguish between these two states with the help of a bit value $b$, as described below.\\
Let us now elaborate on this derivation. The bipartite measures, corresponding to the squared concurrences, in
terms of the parameters of $\ket{\Psi(\boldsymbol{g},z)}$ read
\bea
C_i(\ket{\Psi(\boldsymbol{g},z)}) = \frac{4 r^4 (1- 4 (g_i)^2)(1-16 (g_j)^2(g_k)^2)}{(1+r^4+16r^2 g_i g_j g_k \cos{(2\phi)})^2}. \label{Ci}
\eea
Note that for the sake of simplicity we denote in this proof all bipartite measures with $C_i$ instead of $C_i(\ket{\Psi(\boldsymbol{g},z)})$. Using for instance the equation for $C_2$ we get \begin{small}
\bea \label{Cos2Phi}
\cos(2\phi)\hspace*{-0.2cm} &=&  \hspace*{-0.2cm} \frac{1}{16 g_1 g_2 g_3 r^2} \label{Cos} \\ && \hspace*{-0.2cm}(\frac{2}{\sqrt{C_2}}(\sqrt{(1\hspace*{-0.1cm}-\hspace*{-0.1cm}4 (g_2)^2)(1\hspace*{-0.1cm}-\hspace*{-0.1cm}16 (g_1)^2(g_3)^2)}r^2)\hspace*{-0.1cm}-\hspace*{-0.1cm}1\hspace*{-0.1cm}-\hspace*{-0.1cm}r^4). \nonumber
\eea \end{small}
Note that from Eq.\ \eqref{Cos} we would obtain two solutions for $\phi \in [0,\pi]$, namely $\phi_0$ and $\pi-\phi_0$. However, these two solutions correspond to $\ket{\Psi(\boldsymbol{g},z)}$ and its complex conjugate, which cannot be distinguished with the help of operational entanglement measures \cite{BKLUpaper}, being calculated by just considering the state itself.
Inserting Eq.\ \eqref{Cos} into the equations for $C_1$ and $C_3$ leads to the following equations
\bea
C_1 &=& \frac{C_2 (1-4 (g_1)^2)(1-16(g_2)^2(g_3)^2)}{(1-4 (g_2)^2)(1-16(g_1)^2(g_3)^2)}, \\ C_3 &=& \frac{C_2 (1-4 (g_3)^2)(1-16(g_1)^2(g_2)^2)}{(1-4
(g_2)^2)(1-16(g_1)^2(g_3)^2)}.
\eea
Solving these two equations for  $g_2$ and  $g_3$, using that $g_2, g_3 \geq 0$, leads to solutions in terms of the third parameter $g_1$ given by
\bea
g_2 &=&  \frac{1}{2} \sqrt{\frac{C_1 - C_2 + 4 (g_1)^2 C_3}{4 (g_1)^2 (C_1 - C_2) + C_3}}, \label{g2} \\
g_3 &=& \frac{1}{2} \sqrt{\frac{C_1 - C_3 + 4 (g_1)^2 C_2}{4 (g_1)^2 (C_1 - C_3) + C_2}}. \label{g3}
\eea
Inserting these solutions for $g_2$ and $g_3$ into the equation for the accessible volume, i.e.
\bea
E_a(\ket{\Psi(\boldsymbol{g},z)}) = 8 (1/2 - g_1)(1/2-g_2)(1/2-g_3),
\eea
results in an equation for the third parameter $g_1$. Thus, we have to prove, that this equation has a unique solution for $g_1$. We do so by showing that the first derivative of $E_a$ is always negative in terms of $g_1$. This assures
the existence of a unique solution for $g_1$. The first derivative of $E_a$ in terms of $g_1$ reads (with $g_2, g_3$ given by
Eq.\ \eqref{g2},\eqref{g3}) \begin{small}
\bea
E_a'(g_1) &=& \frac{4 g_1(1-2g_1)(1/2 -g_3)(C_1^2 + C_2^2 - C_3^2-2C_1 C_2)}{(4 (g_1)^2(C_1-C_2) + C_3)^2 g_2}+ \nonumber \\ && [-8 (1/2 -g_2)(1/2 -g_3)] +
\label{Vaderiv} \\ && \frac{4g_1(1-2g_1)(1/2-g_2)(C_1^2 + C_3^2 - C_2^2-2C_1 C_3)}{(4 (g_1)^2(C_1-C_3) + C_2)^2 g_3}. \nonumber
\eea \end{small}
Inserting the expressions for the $C_i$'s from Eq.\ \eqref{Ci}, one can easily show that all three summands in Eq.\ \eqref{Vaderiv} are negative for all valid values of $g_1$, i.e. $\forall g_i \in (0,1/2)$. Hence, the accessible volume, $E_a(g_1)$, is strictly decreasing in $g_1$. Up to now we could show that there exists a unique solution for the three $g_i$ parameters using only the three bipartite measures and the accessible volume. It remains to prove that the parameters $r$ and $\phi$ can be uniquely determined via the source volume, $E_s(r)$, and an additional bit value. As mentioned before, there exist at most two generic states which are compatible with all five entanglement measures in Eq.\ \eqref{Measures}. This is due to the fact that the source volume in Eq.\ \eqref{VSGHZ1} is a monotonous function in terms of $f_z = 2 r^2 \abs{\cos(2\phi)}/(1+r^4)$ for fixed values of the $g_i$'s (they are already uniquely defined). Hence, two states $\ket{\Psi(\boldsymbol{g},z_1)}$ and $\ket{\Psi(\boldsymbol{g},z_2)}$ with the same $g_i$ parameters have the same source volume iff $f_{z_1} = f_{z_2}$. Plugging in the solutions for $\cos(2\phi_1)$ and $\cos(2\phi_2)$ from Eq.\ \eqref{Cos} in terms of $r_1$ and $r_2$, respectively, in the equation $f_{z_1} = f_{z_2}$ and solving for $r_2$ leads to
\bea
r_2 = \sqrt{\frac{1-\sqrt{1-4 u^2}}{2u}}, \label{r2}
\eea
with $u = \sqrt{\frac{C_2}{(1-4 g_2^2)(1-16 g_1^2 g_3^2)}}- r_1^2/(1+r_1^4)$ and for $u\neq0$. Note that if $u=0$, there exists no solution for $r_2$ and $\phi_2$. The corresponding $\phi_2$ parameter is given by Eq.\ \eqref{Cos}, where we simply plug in the expression for $r_2$ in terms of $r_1$. We get a valid solution for $\phi_2$ if the right hand side of Eq.\ \eqref{Cos} is within the range of $\cos(2\phi_2)$. Hence, there exist at most two states that are compatible with all five entanglement measures (in case there exists a valid solution for $r_2 \in (0,1]$ and $\phi_2 \in [0,\pi]$). Note that the sign of $\cos(2\phi)$ for the two states is always different, i.e. $\text{sign}(\cos(2\phi_1))=-\text{sign}(\cos(2\phi_2))$. Furthermore, we can exclude the case where $\cos(2\phi_1)=0$ (i.e. $u=0$), as in this case no second state with parameters $r_2$ and $\phi_2$ exists, that would be compatible to the same five entanglement measures. The source set and the accessible set of the two states are always disjoint, i.e. $M_s(\ket{\Psi(\boldsymbol{g},z_1)}) \bigcap M_s(\ket{\Psi(\boldsymbol{g},z_2)}) = \emptyset$ and $M_a(\ket{\Psi(\boldsymbol{g},z_1)}) \bigcap M_a(\ket{\Psi(\boldsymbol{g},z_2)}) = \emptyset$, even though the density of states in the sets is exactly the same. This follows from cond.\ \eqref{AEq_GHZdet} and the different sign of $\cos(2\phi)$ for the two states. As one can easily see from the first equation in cond.\ \eqref{AEq_GHZdet}, states with a negative (positive) $\cos(2\phi)$ can only reach and access states having also a negative (positive) $\cos(2\phi')$, respectively. Moreover, states with a negative $\cos(2\phi)$ can only be reached by states in the MES with $z=i$, whereas states with a positive $\cos(2\phi)$ can only be reached by states in the MES with $z=1$. In order to provide a simple way of identifying the state uniquely, we use now the following notation. Let $\ket{\Psi(\boldsymbol{g},z)}$ correspond to the five entanglement measures $ \{C_1(\ket{\Psi}), C_2(\ket{\Psi}), C_3(\ket{\Psi}), E_s(\ket{\Psi}), E_a(\ket{\Psi})\}$, such that there are two solutions compatible with these entanglement measures, namely $\ket{\Psi_1}=\ket{\Psi(\boldsymbol{g},z_1)}$ and $\ket{\Psi_2}=\ket{\Psi(\boldsymbol{g},z_2)}$. Here, either $z_1=z$ and $z_2$ such that $f_{z_2}=f_z$ or $z_2=z$ and $z_1$ such that $f_{z_1}=f_z$. Without loss of generality we assume that $z_1$ is such that $\cos(2\phi_1)>0$. We define the states $\ket{\Phi_{\Psi_1}}=\ket{\Psi(\boldsymbol{h},1)}$ and $\ket{\Phi_{\Psi_2}}=\ket{\Psi(\boldsymbol{h},i)}$, where $\boldsymbol{h} = (g_1,g_2,f_z g_3)$. Note that both states are in the MES and that $\ket{\Phi_{\Psi_1}} \in M_s(\ket{\Psi_1})$ and $\ket{\Phi_{\Psi_2}} \in M_s(\ket{\Psi_2})$. As these two states can be directly computed from the five entanglement measures and as their bipartite entanglement is always distinct, they can be used to identify the state uniquely, as is shown in the following lemma.

\begin{lemma}

A generic state $\ket{\Psi}\equiv \ket{\Psi(\boldsymbol{g},z)}$ in the GHZ-class is uniquely characterized (up to LUs and complex conjugation) by either
\begin{enumerate}[(i)]
\item the five operational measures $ \{C_1(\ket{\Psi}), C_2(\ket{\Psi}), C_3(\ket{\Psi}), E_s(\ket{\Psi}), E_a(\ket{\Psi})\}$ or
\item the five operational measures $ \{C_1(\ket{\Psi}), C_2(\ket{\Psi}), C_3(\ket{\Psi}), E_s(\ket{\Psi}), E_a(\ket{\Psi})\}$ together with the bit value of \\$b =  \begin{cases} 1 \ \text{if} \ E(\ket{\Phi_\Psi}) = \max\{E(\ket{\Phi_{\Psi_1}}), E(\ket{\Phi_{\Psi_2}})\} \\
0 \ \text{else} \end{cases}$ \\ with $E=\sum_{i=1}^3 C_i $.
\end{enumerate}
\end{lemma}

Note that $E(\ket{\Phi_{\Psi_1}})= E(\ket{\Phi_{\Psi_2}})$ cannot occur as explained below.
Thus, the five entanglement measures and the information of whether the corresponding state in the MES and in the source set of the state has the smaller or larger sum of bipartite entanglement uniquely characterizes the entanglement contained in the state. Note that the two cases can of course be combined to case (ii), as the bit value $b$ is only required if there exists two states that are compatible with the five entanglement measures, which can be easily determined as explained above. Note further that the required bit value could simply give the information about wether $\ket{\Phi_{\Psi_1}} \in M_s(\ket{\Psi})$ or $\ket{\Phi_{\Psi_2}} \in M_s(\ket{\Psi})$.

\begin{proof}
We showed already above that there exist at most two generic states, which are compatible with all five entanglement measures in Eq.\ \eqref{Measures}. For both of these states $\ket{\Psi_1}$ and $\ket{\Psi_2}$ there exists a unique state in the MES and in the source set, i.e. $\ket{\Phi_{\Psi_1}}$ and $\ket{\Phi_{\Psi_2}}$, respectively, with parameters $\boldsymbol{h} = (g_1,g_2, f_z g_3)$. The third parameter $h_3$ is fixed via cond.\ \eqref{AEq_GHZdet}. Hence, $\ket{\Phi_{\Psi_1}}$ and $\ket{\Phi_{\Psi_2}}$ have the same parameters $h_i$ but $z$ is either equal to 1 or to $i$ as explained above. It is easy to see that the bipartite entanglement, i.e. $C_i(\ket{\Phi_{\Psi}}) = (1-4h_i^2)(1-16 h_j^2 h_k^2)/(1+8 h_i h_j h_k \Re(z'^2))^2$, is always larger for the state $\ket{\Phi_{\Psi_2}}$ with $z=i$, as $\Re((z=i)^2)=-1$ and the $h_i$'s are the same for both states. Hence, also the sum of the bipartite entanglement of $\ket{\Phi_{\Psi_2}}$ is larger, i.e.
\bea
\sum_{i=1}^3 C_i(\ket{\Phi_{\Psi_2}}) > \sum_{i=1}^3 C_i(\ket{\Phi_{\Psi_1}}).
\eea
Therefore, it is proven that the five entanglement measures $\{C_1(\ket{\Psi}), C_2(\ket{\Psi}), C_3(\ket{\Psi}), E_s(\ket{\Psi}), E_a(\ket{\Psi})\}$ and the defined bit value uniquely characterize the entanglement contained in the state $\ket{\Psi(\boldsymbol{g},z)}$.
\end{proof}

\subsection{Uniqueness of non-generic states in the GHZ-class}
Let us now show, that also the entanglement contained in non-generic states, that are states with vanishing $g_i$ parameters or states in GHZ-MES, can be uniquely characterized via the two new measures and three bipartite measures. As we will see in all these cases the additional bit value is not required. Due to the different behaviour of non-generic states under LOCC we divide this section into two subsections. The first deals with the unique characterization of the entanglement of states with vanishing parameters (including the GHZ-state) and the second with states in GHZ-MES.

\subsubsection{Uniqueness of states with vanishing parameters in the GHZ-class}
As we have shown in Sec.\ I.B, states with vanishing $g_i$ parameters have different expressions for the source and accessible volume than generic states. The expression and also the dimension of the volumes depend on the number of vanishing parameters. The two new measures are summarized in Sec.\ I.B for all different cases. Here, we show that together with the bipartite measures they characterize states with vanishing parameters uniquely. We have to distinguish again between states fulfilling $r=1$ and states fulfilling $r\neq1$. The accessible volume of states with $r=1$ is always four-dimensional, whereas the accessible volume of states with $r\neq 1$ is three-dimensional. Furthermore, the source volume of states with $r=1$ has always one dimension less than the source volume of states with $r\neq1$.
One can easily show that for all these states with vanishing parameters, the two new measures $E_a$ and $E_s$ together with the three bipartite measures, $\{C_i\}_{i=1}^3$, uniquely characterize them. In fact knowing the dimension of $E_a$ and $E_s$ allows us to characterize the states using solely the bipartite measures. For states with $r \neq 1$ the dimension of $E_a(\ket{\Psi(\boldsymbol{g},r)})$ is always three, whereas the dimension of $E_s(\ket{\Psi(\boldsymbol{g},z)})$ decreases with the number of vanishing parameters, starting from three for one vanishing $g_i$. For states with $r=1$ the accessible volume is always four-dimensional, while the source volume is again decreased by one dimension for every additional $g_i$ being equal to zero, starting from dimension two for one vanishing $g_i$.  Hence, knowing these dimensions, tells us the number of vanishing parameters and whether $r=1$ or $r\neq1$. Furthermore, the order of the bipartite measures, for example $C_1(\ket{\Psi(\boldsymbol{g},z)}) > C_2(\ket{\Psi(\boldsymbol{g},z)})$ and $C_1(\ket{\Psi(\boldsymbol{g},z)}) > C_3(\ket{\Psi(\boldsymbol{g},z)})$, reveals which of the three $g_i$'s are
equal to zero. Using this information and plugging in the vanishing parameters into the expression of the three bipartite measures $\{C_i(\ket{\Psi(\boldsymbol{g},z)})\}_{i=1}^3$ given in Eq.\ \eqref{Ci}, allows one to easily solve these equations for the remaining parameters.

\subsubsection{Uniqueness of states in GHZ-MES}
In this section we investigate the characterization of the entanglement of states in GHZ-MES.  As mentioned in the main text, a state $\ket{\Psi(\boldsymbol{g},z)}$ is in this set iff $z= 1$ or $z= i$ and non of the $g_i$'s vanishes, or it is the GHZ-state, that was already discussed in the previous section. The condition $z= 1 $ or $z= i$ is equivalent to $z= 1$ and allowing for negative
values of $g_3$, i.e. $g_3 \in (-1/2, 1/2)$. We will use the latter condition for the following calculations.
As we will show here, the entanglement of a state in GHZ-MES is also characterized by the five entanglement measures in Eq.\ \eqref{Measures}. However, as the GHZ-MES constitutes a three parameter family, a more economical characterization is possible. In particular, we will show that the three bipartite measures and one additional bit, that can be easily obtained from $E_a$, is sufficient.  The reason for that is, that in some instances there exist two states $\ket{\Psi_1}$ and $\ket{\Psi_2}$, which are both compatible with the bipartite measures. The additional bit is then obtained via comparing the accessible volume of these two states $\ket{\Psi_1}$ and $\ket{\Psi_2}$. In the
following we will discuss this result in more detail.

In the main text we use the squared concurrences, i.e. $C_i(\ket{\Psi(\boldsymbol{g},z)}) = 2(1-\tr(\rho_i^2))$, as bipartite measures. Here we use, for the sake of simplicity, two times the minimum eigenvalues of the reduced states, i.e. $E_i^{min} = 2 \lambda_{min}(\rho_i)$, with $\rho_i = \tr_{jk}(\rho_{MES}), \ j \neq k \neq i$, as bipartite measures. This is always possible, as for two-dimensional systems, i.e. qubit systems, the entanglement measures are all monotonous functions of each other.
The bipartite entanglement measure in terms of the parameters $g_i$ is given by
\bea
E_i^{+/-} = \frac{(1 \pm 2 g_i)(1 \pm 4 g_j g_k)}{1 + 8 g_i g_j g_k}, \label{EVMES}
\eea
with $i \neq j \neq k$. Depending on the parameters either $E_i^+$ or $E_i^-$ is the minimum eigenvalue, which we denote by $E_i^{min}$. Our aim is now to prove the following.
\begin{lemma} There exist at most two states in GHZ-MES, $\ket{\Psi_1}$ and $\ket{\Psi_2}$, which are compatible with the three bipartite measures $\{E_1^{min}, E_2^{min}, E_3^{min}\}$ (or $\{C_1, C_2, C_3\}$). Hence, for all states $\ket{\Psi}$ in GHZ-MES either
\begin{enumerate}[(i)]
\item $\{E_1^{min}, E_2^{min}, E_3^{min}\}$  uniquely characterizes the entanglement contained in $\ket{\Psi}$, or
\item together with $\{E_1^{min}, E_2^{min}, E_3^{min}\}$  the bit value of $b =  \begin{cases} 1 \ \text{if} \ E_a(\ket{\Psi}) = \max( E_a(\ket{\Psi_1}), E_a(\ket{\Psi_2}) ) \\
0 \ else \end{cases}$ can then be used to uniquely characterize the entanglement of $\ket{\Psi}$.
\end{enumerate}
\end{lemma}
Equivalently to before the two cases can of course be combined in case (ii).
Hence, given the three bipartite measures and the information of whether the state has the larger or smaller accessible volume of the (at most) two states, which are compatible with the bipartite measures, the state is uniquely defined.

\begin{proof}
As mentioned before, depending on the parameter range of the triple ($g_1,g_2,g_3$), either $E_i^{+}$ or $E_i^{-}$ for $i\in \{1,2,3\}$ corresponds to the minimum eigenvalue. More precisely, we have to distinguish the following cases \footnote{Note that those four cases include all possible combinations of the minimum eigenvalues as $E_3^-= E_3^{min}$ implies that $E_1^-=E_1^{min}$ and $E_2^-=E_2^{min}$.}:
\begin{enumerate}[(1)]
\item For positive $g_i$'s or for $-\frac{1}{2} < g_3 <0 \wedge -g_3 < g_2 <\frac{1}{2} \wedge -\frac{g_3}{2 g_2} < g_1 < \frac{1}{2}$ the minimum eigenvalues
    are
    \bea
    \{E_1^-, E_2^-, E_3^-\}.
    \eea
\item For $-\frac{1}{2} < g_3 <0 \wedge [( 0 < g_2 \leq -g_3 \wedge - 2 g_2 g_3 < g_1 < - \frac{g_2}{2 g_3} ) \vee (-g_3 < g_2 < \frac{1}{2} \wedge -2 g_2 g_3 <
    g_1 < - \frac{g_3}{2 g_2})]$ the minimum eigenvalues are
    \bea
    \{E_1^-, E_2^-, E_3^+\}.
    \eea
\item For $  -\frac{1}{2} < g_3 <0 \wedge 0 < g_2 < -g_3 \wedge -\frac{g_2}{2 g_3} < g_1 < \frac{1}{2}$ the minimum eigenvalues are
    \bea
    \{E_1^-, E_2^+, E_3^+\}.
    \eea
\item  For $  -\frac{1}{2} < g_3 <0 \wedge 0< g_2 <1/2 \wedge 0 < g_1 < -2 g_2 g_3$ the minimum eigenvalues are
    \bea
    \{E_1^+, E_2^-, E_3^+\}.
    \eea
\end{enumerate}
The parameter ranges presented above can be easily mapped to parameter ranges of the triple ($E_1^{min}, E_2^{min}, E_3^{min}$). For instance, $(E_1^-, E_2^-, E_3^+)= (E_1^{min}, E_2^{min}, E_3^{min})$ only if $(E_1^{min}, E_2^{min}, E_3^{min})\in J_2= \{(E_1^{min}, E_2^{min}, E_3^{min}): E_1^{min} + E_2^{min} + E_3^{min} > 2\}$. Similar parameter ranges for all the other cases can be found. It can be easily seen that the parameter ranges $J_1,J_3,J_4$ do not intersect. However, each of them intersects with $J_2$. If the given triple ($E_1^{min}, E_2^{min}, E_3^{min}$) is uniquely in one of the parameter ranges given above, i.e. if either $(E_1^{min}, E_2^{min}, E_3^{min}) \in J_k \diagdown (J_k\bigcap J_2)$, for some $k\in \{1,3,4\}$, or $(E_1^{min}, E_2^{min}, E_3^{min}) \in J_2 \diagdown (J_1 \bigcup J_3\bigcup J_4)$, then Eq.\ \eqref{EVMES} can be easily inverted. Hence, one obtains a unique solution for $g_i$, $\forall i \in \{1,2,3\}$ and therefore a unique state which is compatible with ($E_1^{min}, E_2^{min}, E_3^{min}$).
Otherwise, $(E_1^{min}, E_2^{min}, E_3^{min}) \in J_k\bigcap J_2$, for some $k\in \{1,3,4\}$. Let us assume without loss of generality that $k=1$. In this case one obtains one solution for $(g_1,g_2,g_3)$ by solving $(E_1^{min}, E_2^{min}, E_3^{min})=(E_1^-, E_2^-, E_3^-)$ and another solution is given by the solution of $(E_1^{min}, E_2^{min}, E_3^{min})=(E_1^-, E_2^-, E_3^+)$ \footnote{Note that one of the two possible solutions always corresponds to case (2).}. Let us denote the corresponding states by $\ket{\Psi_1}$ and $\ket{\Psi_2}$.
Hence, in this case the bipartite measures are not sufficient to determine the state uniquely. However, as we will show next, it can be easily decided which of the two states corresponds to the state of interest using the accessible volume. The reason for that is that $E_a(\ket{\Psi_1}) \neq E_a(\ket{\Psi_2})$ for any possible two solutions $\ket{\Psi_1} \neq \ket{\Psi_2}$, which we show now by comparing the solutions for $(g_1,g_2,g_3)$ for the two states $\ket{\Psi_1}$ and $\ket{\Psi_2}$. To be more precise, recall that the accessible volume is given by the following integral (see also Eq.\ (8) from the main text)
 \begin{small}
    \bea
    V_a(\ket{\Psi_{MES}})=  \int_{g_1}^{1/2}\hspace{-0.2cm}\int_{g_2}^{1/2}\hspace{-0.2cm}\int_{\abs{g_3}}^{1/2} \hspace{-0.2cm}\int_{\sqrt{\frac{H}{G} - \sqrt{\left(\frac{H}{G}\right)^2-1}}}^1 dr dh_3 dh_2 dh_1 , \label{AVaMES}
    \eea \end{small}
    with $G=g_1 g_2 \abs{g_3}, \ H = h_1 h_2 h_3$.
We can choose without loss of generality $E_i^{min} \neq 1$ $\forall i \in \{1,2,3\}$ for the following reason. If all three minimum eigenvalues are one, the state is the GHZ-state. The case where two are equal to one and the third is different from one can be excluded, as then two parameters $g_i$ are equal to zero, which does not correspond to a state in the MES. If exactly one of the bipartite measures is equal to one, one obtains two valid solutions, but they are the same, i.e. $\ket{\Psi_1} = \ket{\Psi_2}$. For all other valid parameter ranges of $J_1 \bigcap J_2$ one can show that the absolute value of the $g_i^1$ parameters of $\ket{\Psi_1}$ is always smaller than the absolute value of the $g_i^2$ parameters of $\ket{\Psi_2}$, i.e.
\bea
\abs{g_i^1} < \abs{g_i^2} \ \ \forall i \in \{1,2,3\}.
\eea
Moreover, $\sqrt{\frac{H}{G} - \sqrt{\left(\frac{H}{G}\right)^2-1}}$ is a monotonically increasing function of $G$.
 Then, one can show that for the accessible volume in Eq.\ \eqref{AVaMES} we integrate always over a larger volume for $\ket{\Psi_1}$ than for $\ket{\Psi_2}$, as all the lower limits in the integral are smaller for $\ket{\Psi_1}$. Therefore, the accessible volume of $\ket{\Psi_1}$ is always larger than the one for $\ket{\Psi_2}$. One can show exactly the same if $\ket{\Psi_1}$ corresponds to the solutions of $(g_1,g_2,g_3)$ obtained by solving $(E_1^{min},E_2^{min},E_3^{min})=(E_1^-,E_2^+,E_3^+)$ and $(E_1^{min},E_2^{min},E_3^{min})=(E_1^+,E_2^-,E_3^+)$, respectively. In all cases the accessible volume of the state $\ket{\Psi_2}$ is always smaller.
 Thus, the defined bit value together with the three bipartite measures uniquely defines a state in GHZ-MES, which completes the proof.
\end{proof}

\end{document}